\documentclass[12pt]{article}
\usepackage{amsthm}
\usepackage{color}
\usepackage{mathtools}
\usepackage[latin1]{inputenc}
\newtheorem{definition}{Definition}
\usepackage{amsmath}
\usepackage{amssymb}
\usepackage[latin1]{inputenc}
 \usepackage[bottom]{footmisc}
 \usepackage{pdfpages}
\newtheorem{proposition} {Proposition}
\usepackage{comment}

\newcommand{\bea}{\begin{eqnarray}}
\newcommand{\eea}{\end{eqnarray}}
\newcommand{\be}{\begin{equation}}
\newcommand{\ee}{\end{equation}}

\def\Tr{{\text{Tr}}}

\newtheorem{thm}{Theorem}[section]
 
 \newtheorem{lem}[thm]{Lemma}
 \newtheorem{prop}[thm]{Proposition}
 \theoremstyle{definition}
 \newtheorem{defn}[thm]{Definition}
 \theoremstyle{remark}

 \numberwithin{equation}{section}

\begin{document}

\vskip1cm
\centerline{\bf\large The Jacobian conjecture}

\begin{center}
{\bf  Jacques Magnen}
\end{center}

\centerline {CPHT, Ecole polytechnique,}
\centerline{Institut Polytechnique de Paris}
\centerline{ Ecole Polytechnique,  91120 Palaiseau, France}
\centerline{fjmagnen@gmail.com}

\vskip1cm
{\bf Abstract}: The Jacobian conjecture \cite{Keller} involves the map 
$y= x - V(x)$ where $y, x$ are n-dimensional vectors, $V(x)$ is a symmetric polynomial of degree $d$ %($V_{j}(x)=\sum _{r=2}^{d} V^{r}_{j; x_{k_{1},..., x_{k_{r}}}}$) 
for which the Jacobian  hypothesis holds: $\ \ e^{\Tr \ln(1- V'(x))} =1,\ \forall x$. 
The conjecture states that the inverse map ($x$ as a function of $y$) is also polynomial. The proof is inspired by  perturbative field theory.
We express the inverse map $F(y)= y+ V(F(y))$ as a perturbative expansion which is a sum of partially ordered connected trees. %The  indices  of the $V$'s  are the edges of the trees ; the edges are connected by $V$'s.
%In $V^{r}_{j; x_{k_{1},..., x_{k_{r}}}}$ the edges of indices $j , k_{\ell }$ are tree ordered  :  $(edge\ of\ index\ \ j)<_{tree} (edge\ of\ index\ \ k_{\ell }) \ \ for\ all\ \ell \le r $.

We use the property : $\frac{d F_{k}}{dy_{k}}= (\frac{1}{1-V'(F)})_{k,k} =1+ \sum _{q\ge 1} \frac{1}{q} (Tr (V'(F))^{q})_{with\ q\ edges\ of\ index\ k}$ to extract 
inductively in the index $k$ all the sub traces in the expansion of the inverse map.

 We obtain
$F= F(|\le n)\ \ e^{- Tr \ln(1-V'(F(|\le n)))}$

% where $F(|\le n)$ is the sum of the values associated with all the trees such that :
%- there is no  two  distinct  ordered edges of index one

  %- for $1< k\le n$ if there are two  distinct  ordered edges of index $k$ then "between" them there is an edge of index $< k$.

By the Jacobian  hypothesis $e^{- Tr \ln(1-V'(F(|\le n)))} =1$ and a straightforward graphical argument gives that $degree \ in\ y \ of \ F(|\le n)\le d^{2^{n} -2}$
  \vskip1cm 

\section{Introduction} 

The only reference used here is \cite{malek}\footnote{Further references are given in \cite{sazonov}.}. %(\cite{bass})
The Jacobian conjecture involves the map 
\bea\label{map}
y&=& x - V(x) ,\ \ \ V(0)=V'(0) =0 , \\
&&\mbox{ {\bf where the Jacobian  hypothesis holds: }$\ e^{-\Tr \ln(1- V'(x))} =1\ \forall x$ }. \label{jacob-cond}
\eea
 $x, y$ are vectors with $n$ components and $V(x)$ is also a vector and each component $V_{i}$ is a symmetric polynomial of degree $d>1$ such that $V(0)=V'(0)=0$:
\begin{align}
V_{i}(x)= \sum _{Q=1}^{d}V_{i;j_{1},...,j_{Q}} x_{j_{1}}...x_{j_{Q}}\hskip8cm\\
i\ \mbox{is the incoming index,\ \ \ and the }\ \ j_{1},...j_{Q}\ \mbox{are the outgoing indices of the vertex}
\end{align}

 $V$ is called a vertex.
The inverse map has a perturbative expansion :
\bea\label{FP}
F(y)&=&y+V(F(y)),\\
 \label{FP2}
 F_{i}(y)&=& y_{i}+V_{i}\Big(y+V(F(y))\Big),
 \quad \mbox{where $i$ is called the root},
 \\ \label{pert1}
 &=& y_{i}+V_{i}(y+V(y+V(y+V(y+V(...))))).
 \eea
Following \cite{malek} we have : 
 \begin{lem}
For $V$ satisfying (\ref{map}) and $F_{i}$ given by (\ref{FP}),
 $F_{i_{root}}(y)$ is analytic in $y$ for $|y|< \frac{1}{(2n)^{d}\  |1+  ||V||_{\infty,1} |  }$, \ with\ $ ||V||_{\infty,1}= sup _{indices }|V_{indices}|$.
 \end{lem}
\begin{proof}
 
In  (\ref{pert1}) each term of degree $N$ in $y$ contains at most $N-1$ vertices because the degree of $V(x)$ is 
$\ge 2$. The expansion gives for each $x$ two terms : $y$ and $V(F)$,
thus  each monomial generates in the expansion  at most $(2^{d})^{\#\{V\}}$ terms. For each $x$ in $V(x)$ there is a sum over the vector index which gives at most a $n^{d}$
terms for each monomial; furthermore let us call the index of the root $i_{root}$.
 The sum of the terms of degree $N$ in $F_{i}$ is thus bounded by 
\begin{align}
 \Big( (2n)^{d} sup _{indices }|V_{indices}|\Big)^{N-1} |y|^{N},
\end{align}
therefore the lemma follows.
\end{proof}
 
\begin{thm}\label{jacobAA} (Jacobian conjecture)\footnote{A corollary of the theorem is that the conjecture is also true if we relax the condition $V'=0$; this is explained in \cite{malek} and proved in   the appendix  of this paper.}

Let $F(y) $ given by (\ref{pert1}), (\ref{FP}) where the $ y$'s are  n-dimensional 
and $V=\sum_{Q=2}^{d} V^{[Q]}$ where each $V^{[Q]}$ is a symmetric monomial of degree $Q$ such that the Jacobian hypothesis (\ref{jacob-cond}) holds, 
then $F$ is a polynomial in $y$   of degree 
 \be
D(n, d)\le\ d^{2^{n}-2} .
\ee
\end{thm} 

This article is devoted to the  proof of this theorem. 

\vskip1cm
\noindent
{\bf Aknowledgments} Claude de Calan,  Vincent Rivasseau and myself were introduced to this conjecture by Abdelmalek Abdesselam in 2001-2002.  We tried in the following years to prove it starting from the perturbative field theory approach proposed by  Abdesselam \cite{malek}.  
I am deeply indebted to Vincent Rivasseau for all the old attempts that we considered, discussed.
\vskip0.3cm
\section{Trees}
\vskip0.3cm

\begin{definition}
Each term of the perturbation expansion (\ref{pert1}) corresponds to a tree $S$. A  tree is made  vertices $v$  and edges and $y$'s.  Let $r$ be the number of vertices  and $N$ the number of $y$'s of $S$.
 
 A vertex ${v}$ is given by  ${\cal V}_{v}=\sum_{Q=1}^{d}
 V_{\bar{i}_{v}};j_{1,v},\cdots ,j_{Q,v}$; 
 the set of the vertices of $S$ is
 \be \cup_{v\in S}[\sum _{Q=1}^{d}V_{\bar{i}_{v};j_{1,v},...,j_{Q,v}} ],
 \ee
 and the indices of the $y$'s are $\bar{i}_{y}, y\in S$.
 
 Each graph corresponds to a pairing of one index of $\{i_{root},\cup_{v}(j_{1,v},...,j_{Q,v})\}$ (set of the outgoing edges) with an index of $\{\bar{i}_{v}\}\cup\{\bar{i}_{y}, y\in S\}$ (set of the incoming edges).
  $i_{root}$ is the index of the root.
   \end{definition}
   
 Each pair corresponds to an edge $\ell $ which pairs an outgoing edge $\ell _{out}$ with  incoming edge $\ell _{in}$. 
In the perturbative expansion each new step correspond to the dots in (\ref{pert1}). An outgoing edge of the first $V$ is the incoming edge of one of the other $V$ or to a $y$.
Thus in each term of the expansion  all the vertices are connected.
 
The value $F_{S}$ of a graph with $r$ vertices and $N$ $y$'s is
\begin{align}
F_{S} = \frac{1}{r! N!} \sum _{\{\ell _{out}, \ell \in S\}}\Big(\prod _{\ell \in S} \delta_{\ell _{out},\ \ell _{in}}\Big) \prod _{v\in S}{\cal V}_{v} .
\end{align}
The inverse factorials come from the sum over all the permutations of the $v$'s and the $y$'s.

\begin{definition}
%\label{length}
$V'_{i_{1},i_{2}} (F)= \frac{\delta}{\delta F_{i_{2}}}   V_{i_{1}}(F)$.

A subdiagram : $V'_{i_{1},i_{2}}(F)   V'_{i_{2},i_{3}}(F)...V'_{i_{p},i_{p+1}}       $ is a subtree of aligned edges between the $V'$'s and $p$ is the length of this subtree.

If $i_{1}$ corresponds to the edge $\ell $ and $i_{p+1}$ corresponds to the edge $\ell'$  we write it also as $(\frac{1}{1-V'(F)})_{\ell ,\ell '}$.
\end{definition}

\begin{definition}  
{\bf The partial  tree order $<_{tree}$}:

It is defined on each  $F_{S}$ given by (\ref{pert1}).

The $y$'s are maximal in the tree order.
 
If $\ell $ is the incoming edge of a vertex and $\ell' $ is an outgoing edge of the same vertex, then $\ell <_{tree } \ell '$. 

If   ${\ell } <_{tree} \ell'  $ then the two edges are aligned.

If   ${\ell } <_{tree} \ell' <_{tree} \ell'' $ then $\ell ' $ is (by definition) 
\emph{between}
 %$_{<_{tree}}$\ \  
 $\ell $ and $\ell "$.
\end{definition}
 
\begin{definition}  
{\bf The complete  around the   tree order $<_{around\ tree}$}:

If $\ell <_{tree}\ell '$ then by definition  $\ell <_{around\ tree} \ell '$.

In $S$ for each vertex $(V^{[Q]})_{i; j_{1,v},...,j_{Q,v}} $ if $\ell $ and $\ell '$ are two outgoing edges of $v$ 
of outgoing index respectively $j_{k,v}$ and  $j_{k',v}$ and if $k<k'$ then
$\ell <_{around\ tree} \ell '$.
\end{definition}
Then by associativity the around the tree order is a complete order.

The proof of the theorem (\ref{jacobAA}) is made in successive steps 
labelled by $k=1,..., n$.

%\vskip1cm {\bf III)
\section{The first step $k=1$}

\begin{proposition}
%\label{color1}
\begin{align}
F_{i}= F_{i}(|1) \ e^{-\Tr\ln(1-V'(F(|1)))}
\end{align}
where $F(|1) $ means the sum of graphs such that there is no aligned edges of index one. 
\end{proposition}
\begin{proof}

We introduce a perturbation variable $s^{1}$\  and replace each (\ref{FP}) by
\bea\label{s1}F(s^{1}) = f_{s^{1}}.y+ f_{s^{1}}.V(F(s^{1}))\quad
with\ \ (f_{s^{1}})_{\alpha }=1+\delta_{\alpha ,1}(s^{1}-1). 
\eea
Then we expand :
\bea\label{1,1a}
F_{i}(1)= F_{i}(0)+ \int _{0}^{1} ds^{1} \frac{d}{ds^{1}}F_{i}(s^{1})=\sum _{p^{1}} \frac{1}{p^{1}!}\Big[ \frac{d^{p^{1}}}{(ds^{1})^{p^{1}}}F_{i}(s^{1})\Big]_{s^{1}=0}.
\eea
Deriving (\ref{s1}) :
\begin{align}
F'(s^{1}) = f'_{s^{1}}.y+ f'_{s^{1}}.V(F(s^{1})) +f_{s^{1}}.[V'(F(s^{1}))].F'(s^{1})\\
(1-f_{s^{1}}.V'(F(s^{1})))F'(s^{1}) = y_{1}+V_{1}(F(s^{1}))\\
F'(s^{1}) =\Big[ \frac{1}{1-f_{s^{1}}.V'(F(s^{1})}\Big]_{1} (    y_{1}+V_{1}(F(s^{1}))\\
more\ precisely\ \  F_{j}'(s^{1}) =\Big[ \frac{1}{1-f_{s^{1}}.V'(F(s^{1})}\Big]_{j,1} (    y_{1}+V_{1}(F(s^{1}))\
\end{align}

Let $S^{1}$ be the set of all the edges of index one in $S$ and let $p^{1}$ be the number of these edges.
\begin{align}
F_{i}(1)= \sum _{S^ {1}}  \frac{1}{p^{1}!}   \frac{d^{p^{1}}}{(ds^{1})^{p^{1}}} F_{i}(s^{1}).
\end{align}
This means that $ \frac{d^{p^{1}}}{(ds^{1})^{p^{1}}}F_{i}(s^{1})$ is not $s^{1}$ dependent. 

We give an order to the $p^{1}$ derivatives and let  $\ell (r)$ be the edge derived by the $r^{th}$ derivative. 
The result of  $ \frac{d^{p^{1}}}{(ds^{1})^{p^{1}}}F_{i}(s^{1})$ is a sum of terms corresponding to all the $\{\ell (r)\}$
which runs over all the  permutations of the $p^{1}$ elements of $S^{1}$.

Then we define:
\begin{defn}
\begin{align}
\ell _{1}^{1}= inf_{<around\ the\ tree}\{\ell \in S^{1}\},\\
\ell _{j,sup}^{1}= max_{<tree}\{\ell \in S^{1}, \ell >_{tree} \ell _{j}^{1}\}\ \ j\ge 1,\\
if \  max_{<tree}\{\ell \in S^{1}, \ell >_{tree} \ell _{j}^{1}\}=\emptyset, \ \ then \ \ell _{j,sup}^{1}=\ell _{j}^{1},\\ 
\ell _{j}^{1}= inf_{<around\ the\ tree}\{\ell \in S^{1},\ \ell >_{around\ tree} \ell _{j-1,sup}^{1}\} .
\end{align}
Let $k_{j}= \#\{\ell \ni S^{1},\ \ \ \ell _{j}^{1}<_{tree}\ell \le_{tree}\ell _{j,sup}^{1}\}$ and $k_{j}=0$ if $\ell _{j}^{1}=\ell _{j,sup}^{1}$.
\end{defn}
\vskip0.6cm
\begin{definition} 
For $j$ s.t.\  $\ell^{1}_{j}<_{tree}\ell^{1}_{j,sup}$\ let :
\ \ \ $R^{1}_{j}=   \Big(\frac{1}{1-V'(F(0))}\Big)_{\ell^{1}_{j},\ell^{1}_{j,sup}}$\ ,

 \ and  for $\ell _{j}^{1}=\ell _{j,sup}^{1}$ \ \ \ \ \ \ \  \ $R^{1}_{j}=1$.

 Let \  $P^{1}=\#\{j \ \ s.t. \ \   \ell ^{1}_{j}\ exists\}$\ \  be the number of resolvent of step one.

\end{definition} 

A contribution corresponding to $p^{1}, P^{1}, \{k_{i}\}$ is
\begin{align}
F_{i_{root}} \prod _{j=1}^{P^{1}}  \Big(\frac{1}{1-V'(F(0))}\Big)_{\ell^{1}_{j},\ell^{1}_{j,sup}} F_{\ell^{1}_{j,sup}}(0).
\end{align}

Summing over all the orders of the derived edges corresponding to $k_{j}$ e.g all the orders $\{\ell \ni S^{1},\ \ \ \ell _{j}^{1}\le_{tree}\ell \le_{tree}\ell _{j,sup}^{1}\}$ \ gives:
\bea
\label{factor1}
(k_{j}+1)!\ \  \Big((\frac{1}{1-V'(F(0))})_{1,1}\Big)^{k_{j}}.
\eea
The number of configuration of the edges of $S^{1}$ corresponding to $p^{1}, P^{1},\ {k_{j}}$ is 
\bea\label{factor2}
\frac{p^{1}!}{P^{1}!  \prod _{j} (k_{j}+1)!}\; .
\eea
 Thus the factorials in $ \prod _{j} (k_{j}+1)!$ are cancelled between (\ref{factor1}) and (\ref{factor2}).
 
For each $j$ it remains: 
\bea\label{trace}
\Big(\frac{1}{1-V'(F(0))}\Big)_{\ell^{1}_{j},\ell^{1}_{j,sup}}=\Big[\Big(\frac{1}{1-V'(F(0))}\Big)_{1,1}\Big]_{\substack{with\ k_{j}\ vertices\ V'\\ with\ at\ least\ one\  index\  one}}\cr
  = \frac{1}{k_{j}} \Tr \frac{1}{1-V'(F(0))}\Big|_{k_{j} \ edges\ of\ index\ one\ in\ the\ trace},
  \eea
and summing over $k_{j}$ we get
  \bea
  -\Tr \ln(1-V'(F(0))) \Big|_{at \ least\ one \ edges\ of\ index\ one\ in\ the\ trace}\ .
\eea
We introduce $F(|1)$:
\begin{definition}
$F(|1)$ is equal to the sum of graphs where there are no edges of index one aligned.
\end{definition}

The result of the expansion of $F_{i}$ is thus
\bea
&&\hskip-1.4cm F_{i}= \sum _{S^{1}}
\Big[ F_{i}(0) \sum _{ P }   \frac{1}{P!}
 \Big(\Tr\ln (\frac{1}{1-V'(F(0))})\Big|_{\substack{trace\ with\\ at\ least\\ one\ edge\ of\\ index\ one} }\Big)^{P}
\prod _{\substack{j\ s.t.\ \ell^{1}_{j},\ \ell^{1}_{j+1}\\ are\  not\ aligned}} F_{\ell^{1}_{j}}(|1)\Big]_{S^{1}},\\ \label{step1}
&&F_{i}=   F_{i}(|1)  e^{\big\{-\Tr\ln (1-V'(F(|1)))\Big|_{trace\ with\ at\ least\ one\ edge\ of\ index\ one} \big\}}.
\eea
\end{proof}

\section{The k$^{th}$ step $i=k$ and the factorisation of all same index alignments}
The induction hypothesis is
\bea
F_{i}= 
 F_{i}(| <k) \prod _{r<k}\Big[ e^{-\Tr\ln (1-V'(F(|<k)))}\Big]_{\substack{traces\ with\ no\ edge\ of\ index\ \le\ r\ and\\  at\ least\ one\ edge\ of \ index\ k}}\ \ .
\eea
\begin{definition}
If two edges $\ell ,\ell ',\ \  \ell <_{tree}\ell '$ of index $q< k$ are aligned$^{k}$ if  
there exists no edge $\ell "$ such that 
$\ell "$ is of index $<q$ and with $\ell <_{tree}\ell "<_{tree}\ell '$.

Where $ F(| <k) $ means $F$ restricted to graphs such that any pair of edges of the same index $q<k$  are  not aligned$^{q}$ (e.g.\ there is no edge of index $<q$ between$_{<_{tree}}$ the two edges of the pair).   
\end{definition}
For each $S$ we use the tree order  and the around the tree order (induced from the orders on $S$)  \ on the set 
$\cup_{F(|<k)}(graph\ corresponding\ to\ F(|<k))$,
and we can define $\ell ^{k}_{j}$ in the same way as the $\ell ^{1}_{j}$ but with ``aligned" replaced by  aligned$^{k}$.

We then obtain in a way analogous as for (\ref{step1})
\bea\label{stepk}
F_{i}= 
 F_{i}(| \le k) \prod _{r\le k}\Big[ e^{-\Tr\ln (1-V'(F(|\le k)))}\Big]_{\substack{traces\ with\ no\ edge\ of\ index\ \le\ r\ and\\  at\ least\ one\ edge\ of\ index\ k}}\ \ .
\eea
Then the following proposition holds
\begin{prop}\label{factorisation}(factorisation finale)

\bea\label{stepn}
%i)\ \ 
F_{i} &= 
 F_{i}(| \le n) \prod _{r\le n}\Big[ e^{-\Tr\ln (1-V'(F(|\le k)))}\Big]_{\substack{traces\ with\ no\ edge\ of\ index\ <\ r\ and\\  at\ least\ one\ edge\ of\ index\ r}}\\
 &= F_{i}(| \le n) \ .
\eea
\end{prop}

\section{Proof of the Jacobian conjecture}

\begin{lem}\label{jacobi=cond}\ \ 
$\prod _{r\le n}\Big[ e^{-\Tr\ln (1-V'(x))}\Big]_{\substack{traces\ with\ no\ edge\ of\ index\ <\ r\ and\\  at\ least\ one\ edge\ of\ index\ r}}= 1$\ .
\end{lem}
\begin{proof}
\begin{align}
\Big[Tr(V'(x))^{Q})\Big]=\Big[Tr(V'(x))^{Q})\Big]_{traces\ with\ at\ least\ one\ edge\ of\ index\ 1}\cr +
\sum_{r=2}^{n} \Big[Tr(V'(x))^{Q})\Big]_{\substack{traces\ with\ no\ edge\ of\ index\ <\ r\ and\\  at\ least\ one\ edge\ of\ index\ r}}\ .
\end{align}
which is equal to one by the Jacobian hypothesis (\ref{jacob-cond}).
\end{proof}

\begin{lem}
The length of each  $F(|\le n)$ 
is bounded by  $ \le 2^{n}-1$.
\end{lem}
\begin{proof}

We proceed inductively on the number $n$ of indices. Let  $L_{n}$ be the maximal length of the trees in  $T(n)$.
For $n=1$ a tree of  $T(1)$  is made of only one edge so  $L_{1}= 1$.
For $n=2$ each edge $\ell _{1}$ of index $1$ splits each tree in 

[a sub-tree containing at most one edge (then of index $2$) ]  $<_{tree} \ell _{1}$ and 

$\{$[sub-trees containing each  at most one  edge (then of index $2$ ) ] $\}$   $>_{tree} \ell _{1}$,

 thus  $L_{2}= 2+1= 2L_{1}+1$.

For each new index $k$ each edge $\ell _{k}$ of index  $k$ splits each sub-tree in sub-sub-tree   either $<_{tree} \ell _{k}$ or sub-sub-trees $>_{tree}\ell  _{k}$ thus
$L_{n} =1+2+...+2^{n-2}+2^{n-1}L_{1}= 2^{n}-1$.

\item We consider the induction hypothesis: ``A tree of length $p$  is at most of degree $d^{p-1}$".

A tree of length $1$ is of degree one.

A tree of length $p$ is at most of degree
$d\times$ the bound on the degree of a tree of length $p-1$
$=d\ d^{p-2}$; thus the induction hypothesis is proved.
%iii) by construction in the union of all the cycles  there is at most one edge of any index. 
\end{proof}

\begin{proof}[Proof of Theorem \ref{jacobAA}]

Let $(degree)_{L}$ be the degree in $y$ of the sum of the  graphs of length $\le L$ .
\begin{lem}\ \ \ 
 $(degree)_{L}\le d^{L-1}$ \ .
\end{lem}
\begin{proof}
Let $F_{L}(y)$ the value of the sum of the  graphs of length $L$. 
We proceed by induction:  $(degree)_{1}=1= d^{1-1}$,
and $F_{L+1}(y)= F_{L}(y+V(y))$, thus $(degree)_{L+1}= d\ (degree)_{L} $. This proves the induction.
\end{proof}

Finally :
\begin{align}
D(n,d) \le  (degree)_{[2^{n}-1]} = d^{2^{n}-2}.
\end{align}
%which proves Theorem 1
\end{proof}

\noindent
{\bf Appendix : Theorem (\ref{jacobAA}) is also true for $V'(0)\ne0$} 
\vskip0.3cm
If $V^{[1]}\ne 0$ then  we write   $\frac{1}{1-V'}$ as\ \ $\frac{1}{1-V'}=\frac{1}{1- (V^{[1]})'-(V-V^{[1]}  )' }= \frac{1}{1- (V^{[1]})'} \frac{1}{1-\frac{ (V-V^{[1]}  )'   }{1- (V^{[1]})'} }$.  
\vskip0.6cm   $F$ is given by the same formula (\ref{pert1}) but where $V'$ is replaced by $ (V-V^{[1]})'$  (where there is no linear term)  and the sum over indices are replaced by convolution with $\frac{1}{1- (V^{[1]})'}$. 
 
 Then the Jacobian conjecture follows from $\frac{ (V-V^{[1]}  )'   }{1- (V^{[1]})'} $ has the  properties which were used to prove Theorem (\ref{jacobAA}):

 \begin{itemize}
 \item
  $\frac{1}{1- (V^{[1]})'}$ is an operator which matrix element are uniformly bounded:
 
$ (V^{[1]})'$ being constant in $y$ then  ($(V^{[1]})'= (V^{[1]})'(0)=V'(0)$) so that  $\det (1-V'(0))=\det (1-(V^[1])')= 1$ which implies using Cayley's theorem that $((V^{[1]})')^{n}=0$, see \cite{malek}.
Thus  $[\frac{1}{1-(V^{[1]})'}]_{i,j}= \sum _{k=0}^{n} [((V^{[1]})')^{k}]_{i,j}\le (n+1) [||(V^{[1]})')||_{\infty,1}]^{n}$ which is  uniformly bounded.

\item
 $(V-V^{[1]}  )' (F)$ being of degree $\ge 1$ in $F$, so that using the previous point, $F$ is a function analytic in $y$ for $y$ small enough.
 
\item
All the terms of degree in $y$ bigger than $n^{2}d^{(2^{n}-1)}$ are zero by the same argument giving the proof of Theorem (\ref{jacobAA}):
the main ingredient was that $\det(1-V'(F))=1,\  \forall \ F$.
 \end{itemize}

\end{document}